\documentclass[fleqn,usenatbib]{mnras}

\usepackage[T1]{fontenc}
\usepackage{newtxtext,newtxmath}
\usepackage{graphicx}
\usepackage{amsmath}
\usepackage{bm}
\usepackage{multicol}

\graphicspath{{image/}}

\hypersetup{colorlinks=true,linkcolor=blue,citecolor=blue,urlcolor=blue,hypertexnames=false}

\newcommand{\dd}{\mathrm{d}}
\newcommand{\Msun}{M_\odot}
\newcommand{\phistar}{\phi_\star}
\newcommand{\phieff}{\phi_{\rm eff}}
\newcommand{\meff}{m_{\rm eff}}
\newcommand{\MBHmax}{M_{\rm BH,max}}
\newcommand{\nhat}{\hat{\bm n}}
\newcommand{\phat}{\hat{\bm p}}
\newcommand{\khat}{\hat{\bm k}}
\newcommand{\uhat}{\hat{\bm u}}
\newcommand{\vhat}{\hat{\bm v}}
\newcommand{\Mc}{\mathcal{M}}
\newcommand{\Neff}{N_{\rm eff}}
\newcommand{\ellM}{\ell_M}

\title[SMBHB population coordinates in PTA inference]{Astrophysical Population Coordinates for Supermassive Black Hole Binaries in Pulsar Timing Array Inference}

\author[Li et al.]{
Yikun Li,$^{1,2}$
Muhammad Ahmad,$^{1,2}$
Shaoguang Guo,$^{3,2}$
and Lang Cui$^{1,2,4}$\thanks{E-mail: cuilang@xao.ac.cn}\\
$^{1}$State Key Laboratory of Radio Astronomy and Technology, Xinjiang Astronomical Observatory, CAS, 150 Science 1-Street, Urumqi 830011, China\\
$^{2}$School of Astronomy and Space Science, University of Chinese Academy of Sciences, No.19A Yuquan Road, Beijing 100049, China\\
$^{3}$Shanghai Astronomical Observatory, Chinese Academy of Sciences, Shanghai 200030, China\\
$^{4}$Xinjiang Key Laboratory of Radio Astrophysics, 150 Science 1-Street, Urumqi 830011, China
}

\date{Accepted XXX. Received YYY; in original form ZZZ}
\pubyear{2026}

\begin{document}
\label{firstpage}
\pagerange{\pageref*{firstpage}--\pageref*{lastpage}}
\maketitle
\begin{abstract}
Pulsar timing arrays can probe the population physics of supermassive black-hole binaries through the nanohertz gravitational-wave background.  We construct a phenomenological forward model that follows source abundance, binary residence time, the high-mass population, finite-source strain moments, and the pulsar timing response.  The simulated observables constrain three standardized population coordinates: $\beta$, which controls the residence-time and spectral response; $\phieff$, which describes source normalization after accounting for its covariance with $\beta$; and $\meff$, which is dominated by the high-mass cutoff.  In the evaluation ensemble, the posterior-mean correlations with the simulated values are $0.928$, $0.926$, and $0.884$, with central 90 per cent coverages of $0.938\pm0.015$, $0.871\pm0.021$, and $0.898\pm0.019$, respectively.  Frequency-resolved observables are most important for $\beta$, and strain moments beyond a common-process power law provide sensitivity to the normalization and high-mass coordinates; the fourth strain moment identifies $\meff$ with rare, massive binaries.  These coordinates quantify the relative sensitivity of the adopted PTA summaries within this population model, for which nearby population realizations retain substantial posterior overlap.
\end{abstract}


\section{Introduction}
\label{sec:introduction}

Pulsar timing arrays (PTAs) measure correlated timing variations from gravitational waves at nanohertz frequencies, where the expected source population is dominated by supermassive black-hole binaries (SMBHBs).  Independent analyses by NANOGrav, the European and Indian PTAs, the Parkes PTA, the Chinese PTA, and the International PTA have established a common red process and increasing evidence for the spatial correlations expected from a gravitational-wave background \citep{NANOGrav:2023gor,EPTA:2023fyk,Reardon:2023gzh,Xu:2023wog,InternationalPulsarTimingArray:2023mzf,Miles:2024seg}.  If SMBHBs provide the dominant contribution, these measurements open a direct route from the nanohertz spectrum to the assembly of massive galaxies and their central black holes.  The scientific objective then extends beyond establishing a background: its amplitude, frequency dependence, source fluctuations, and angular structure can constrain the abundance, masses, redshift distribution, and orbital evolution of the binaries \citep{Rajagopal:1995zz,Jaffe:2003qp,Phinney:2001di,Sesana:2004sp,Sesana:2008mz,BurkeSpolaor:2019xpf}.

The gravitational-wave background combines several stages of SMBHB formation and evolution.  Galaxy merger rates set the supply of potential binaries, black-hole--host relations and mass-ratio distributions determine their masses, and dynamical friction, stellar scattering, gaseous interactions, and gravitational radiation regulate the time spent in each frequency interval.  The high-mass tail is especially influential because the strain rises steeply with chirp mass, even though the corresponding systems are rare.  Redshift evolution changes both the number of sources and the frequencies at which they enter the observed band.  Population calculations therefore connect a measured spectrum to several coupled ingredients of galaxy and black-hole evolution \citep{Sesana:2004sp,Sesana:2008mz,Sesana:2009ss,Ravi:2012bz,Kelley:2016gse,Hazboun:2019vhv,NANOGrav:2023hfp}.  Changes in merger abundance, mass normalization, and binary residence time can compensate one another, so comparable background amplitudes may arise from distinct population histories.

A common-process amplitude and spectral index provide an efficient description of the mean signal and are well matched to present PTA searches.  The spectral index carries information about binary residence time and environmental evolution, and the amplitude measures an integrated combination of source abundance and mass.  Population interpretations based on these quantities already constrain combinations of merger efficiency, black-hole scaling relations, and binary dynamics \citep{Kelley:2016gse,NANOGrav:2023hfp,Sato-Polito:2023gym,Liepold:2024woa,IzquierdoVillalba:2022gcs}.  The same compression also exposes a central inference problem: individual astrophysical parameters often move the amplitude and slope in correlated directions.  Population coordinates aligned with the response of the observables can express the measured combinations directly and retain a clear connection to the underlying astrophysics.

Finite source numbers provide further information about those combinations.  A nanohertz frequency bin can contain a small number of high-strain binaries, producing frequency-to-frequency fluctuations, departures from Gaussian strain statistics, and a distinction between an unresolved background and individually bright sources \citep{Ravi:2012bz,Becsy:2022pnr,Sato-Polito:2024lew,Lamb:2024gbh,Allen:2022dzg,Allen:2024mtn}.  The second strain moment measures the mean power contributed by the population.  The fourth moment weights the brightest sources more strongly and consequently responds sharply to the upper end of the mass distribution.  Ratios of these moments describe the effective number of participating sources and connect stochastic-background measurements to continuous-wave searches.  Sky anisotropy supplies a related view of source discreteness and large-scale structure through pulsar-pair correlations and low-order angular modes \citep{Mingarelli:2013dsa,Mingarelli:2017fbe,Taylor:2013esa,Cornish:2013aba,NANOGrav:2023tcn,Gardiner:2023zzr,Chen:2026mid}.  Frequency-resolved and finite-source observables can therefore separate population responses that remain combined in the mean power law.

Existing population-synthesis studies have mapped galaxy demographics and binary evolution into predictions for the background spectrum, including the range allowed by black-hole scaling relations and environmental hardening.  Studies of finite populations have quantified spectral variance, non-Gaussianity, resolvable binaries, and the influence of rare massive systems.  Angular analyses have developed complementary statistics for anisotropy and the spatial correlation pattern, and recent statistical methods permit population posteriors to be estimated from increasingly detailed simulations.  These developments establish the physical ingredients and observable signatures needed for population inference.  A remaining task is to identify a compact set of population combinations that follows the response of the observables, admits direct physical interpretation, and can be evaluated with the spectral and finite-source summaries available to PTA analyses.

We address this task with a phenomenological SMBHB forward model.  The calculation evaluates source counts on a grid in observed frequency, redshift, black-hole mass, and sky direction; includes a frequency-dependent residence time and a variable high-mass cutoff; and forms the second and fourth strain moments of the resulting population.  Individually bright sources and the unresolved background are propagated through the pulsar-pair response and timing-residual model.  The resulting frequency, moment, and angular summaries are used in simulation-based posterior estimation, which provides a likelihood-free treatment of the stochastic source population \citep{Cranmer:2020sbi,Papamakarios:2019snl,Papamakarios:2021flows,Greenberg:2019apt,TejeroCantero:2020sbi}.  This construction keeps the physical source model, observable summaries, and posterior target in a single calculation.

The analysis identifies three population coordinates, $\Theta_{\rm eff}=(\beta,\phieff,\meff)$.  The residence-time exponent $\beta$ controls the principal spectral response.  The coordinate $\phieff$ describes source normalization after accounting for its covariance with residence time, and $\meff$ follows the high-mass contribution to finite-source strain moments.  We measure their recovery in simulated observations, compare posterior geometry with the local spacing of population realizations, quantify sensitivity to selected observable families, and trace the mass dependence of $\meff$.  Matched three-dimensional calculations relate these coordinates to the conventional $(A,\gamma_{\rm cp},N_{\rm eff})$ description and examine their response under representative population and PTA configurations.  The public NANOGrav 15-year free spectrum supplies an observational comparison in amplitude and spectral-index space; the three-coordinate population posteriors reported here are obtained from simulated observables.

\section{Astrophysical Forward Model}
\label{sec:astrophysical-model}

The forward calculation starts from a phenomenological SMBHB population and evaluates its source counts on a grid in observed frequency, redshift, black-hole mass, and sky direction.  The binned population determines the second and fourth strain moments and the division between individually bright and unresolved sources.  These quantities are projected through the pulsar-pair response and residual model to form the observables used in the population inference.

\subsection{Source-count kernel and discretized source space}
\label{subsec:continuous-to-discrete}

A useful starting point is the continuous expected source distribution $\dd \bar{N}/(\dd\ln f\,\dd z\,\dd\ln M\,\dd\Omega)$, where $f$ is the observed GW frequency, $z$ is redshift, $M$ is the black-hole mass scale used by the model, and $\Omega$ is sky direction.  The discretized expected count in a cell is the corresponding integral,
\begin{align}
    \lambda_{ijkp}
    &\simeq
    \left.
    \frac{\dd \bar{N}}
    {\dd\ln f\,\dd z\,\dd\ln M\,\dd\Omega}
    \right|_{f_i,z_j,M_k,\nhat_p}
    \nonumber\\
    &\quad\times
    \Delta\ln f_i\,\Delta z_j\,\Delta\ln M_k\,\Delta\Omega ,
    \label{eq:lambda-continuous}
\end{align}
or, equivalently, with $\Delta\ln f_i=\Delta f_i/f_i$ absorbed into the residence factor.  In the present phenomenological description, the continuous source density is represented by a cell-count kernel and then normalized to a total expected count.  The equations below define a reference population model in which residence weighting, redshift evolution, high-mass loading, and sky modulation can be traced into PTA observables.

The emitted and observed GW frequencies obey the standard redshift relation $f_r=(1+z)f$.  In this convention, redshift enters the strain through the detector-frame chirp mass and the luminosity distance, and the observed-frequency kernel in Eq.~\eqref{eq:residence-time} represents the residence dependence.  The residence kernel contains the binary-hardening response considered here.

\subsubsection*{Discretized source space.}
\label{subsec:grid}

The forward model is a discretized SMBHB population model on a four-way grid in observed GW frequency $f_i$, redshift $z_j$, black-hole mass scale $M_k$, and sky direction $\nhat_p$.  Frequency and redshift bin widths are denoted by $\Delta f_i$ and $\Delta z_j$, the mass-bin edges by $M_{k,-}$ and $M_{k,+}$, and the pixel area by $\Delta \Omega = 4\pi/N_{\rm pix}$.  Table~\ref{tab:source-grid} summarizes the fiducial source grid.  Frequencies and masses are spaced geometrically, redshifts are spaced linearly, and the logarithmic mass-cell width used in the population kernel is $\Delta \ln M_k=\ln(M_{k,+}/M_{k,-})$.

\begin{table}
    \centering
    \caption{Fiducial source-space grid used by the forward model.}
    \label{tab:source-grid}
    \resizebox{\columnwidth}{!}{%
    \begin{tabular}{lll}
        \hline
        Coordinate & Range & Number of bins \\
        \hline
        observed GW frequency $f_i$ & $1$--$8~{\rm nHz}$ & $N_f=4$ \\
        redshift $z_j$ & $0.1$--$1.2$ & $N_z=3$ \\
        black-hole mass scale $M_k$ & $10^8$--$3\times10^{10}\,\Msun$ & $N_M=4$ \\
        sky direction $\nhat_p$ & full sky & $N_{\rm pix}=192$ \\
        \hline
    \end{tabular}
    }
\end{table}
This matters because the population model is evaluated as an explicit cell-count kernel with traceable residence, redshift, mass, and sky factors.

\subsubsection*{Phenomenological source-count kernel.}
\label{subsec:population-kernel}

The population model assigns an expected number of SMBHBs to every $(f_i,z_j,M_k,\nhat_p)$ cell.  The isotropic kernel before the LSS modulation is an unnormalized positive cell weight: it is separable enough to expose the roles of the population parameters, and it is normalized after all frequency, redshift, mass, and sky factors have been combined \citep{Phinney:2001di,Sesana:2008mz,Kelley:2016gse,NANOGrav:2023hfp,Liepold:2024woa}.  The flat-cosmology distance factors are
\begin{equation}
    d_L(z)=(1+z)\chi(z),
    \qquad
    \frac{\dd V_c}{\dd z\,\dd\Omega}
    =
    \frac{c\,\chi^2(z)}{H(z)}.
    \label{eq:cosmology-volume}
\end{equation}
Here $\chi(z)$ is the comoving distance and $H(z)$ is the Hubble rate.  Therefore the comoving volume in one sky pixel and one redshift bin is $\Delta V_j^{\rm pix}=\left.(\dd V_c/\dd z\,\dd\Omega)\right|_{z_j}\Delta z_j\,\Delta\Omega$.  The unnormalized source-count weight is
\begin{align}
    \mathcal{K}_{ijk}
    &=
    R_z(z_j)\,
    T_f(f_i)\,
    \Delta V_j^{\rm pix}\,
    \Delta f_i\,
    \nonumber\\
    &\quad\times
    \Delta\ln M_k\,
    P_M(M_k)\,
    S_\epsilon(M_k)\,
    \sqrt{D(z_j)} ,
    \label{eq:raw-kernel}
\end{align}
where $D(z)$ is the cosmological growth factor.  The redshift factor is
\begin{equation}
    R_z(z)
    =
    \left[\frac{1+z}{1+z_{\rm min}}\right]^{-\gamma}
    \exp\left(-\frac{z}{z_{\rm turn}}\right),
    \label{eq:redshift-kernel}
\end{equation}
with $z_{\rm turn}=1.6$ in the fiducial configuration.  The mass factor is
\begin{equation}
    P_M(M)
    =
    \left(\frac{M}{10^9\,\Msun}\right)^{-1.15}
    \exp\left[-\left(\frac{M}{\MBHmax}\right)^2\right],
    \label{eq:mass-kernel}
\end{equation}
and the scatter boost is
\begin{equation}
    S_\epsilon(M)
    =
    1+\epsilon_0
    \left[\log_{10}\left(\frac{M}{10^9\,\Msun}\right)\right]^2 .
    \label{eq:scatter-boost}
\end{equation}
The frequency residence factor is
\begin{equation}
    T_f(f)
    =
    \frac{1}{f}
    \left(\frac{f}{f_{\rm min}}\right)^{-\beta}.
    \label{eq:residence-time}
\end{equation}
Equivalently, the kernel enters the count integral as $T_f(f_i)\Delta f_i=(\Delta f_i/f_i)(f_i/f_{\rm min})^{-\beta}$.  Per logarithmic frequency interval, the effective source occupancy therefore scales as $dN/d\ln f \propto (f/f_{\rm min})^{-\beta}$.  The parameter $\beta$ controls the residence weighting across PTA frequencies.  Larger $\beta$ shifts population weight toward the lowest observed frequencies and steepens the strain power response.

The leading physical reference is the circular GW driven inspiral.  For this case $df/dt \propto f^{11/3}$ and hence $dt/d\ln f=f/(df/dt)\propto f^{-8/3}$ \citep[e.g.][]{Phinney:2001di}.  In the convention of Eq.~\eqref{eq:residence-time}, this identifies $\beta_{\rm GW}=8/3$ as the residence-time exponent behind the canonical SMBHB power-law spectrum.  Environmental hardening, selection, and redshift or mass weighting can change the frequency dependence represented by $\beta$ \citep{Begelman:1980vb,Quinlan:1996vp,Cuadra:2009te,2011MNRAS.411.1467K,Sampson:2015ada,Taylor:2017nfs,Lamb:2024gbh}.

The limiting behavior of Eq.~\eqref{eq:raw-kernel} is the intended physical guide.  Increasing $\phistar$ scales the total expected count after normalization.  Increasing $\beta$ gives more weight to lower observed frequencies relative to higher ones.  Increasing $\MBHmax$ weakens the exponential cutoff at the high-mass end.  Increasing $\epsilon_0$ strengthens the mass-dependent scatter boost away from the pivot mass $10^9\,\Msun$.

The source-count normalization is global over all sky pixels:
\begin{equation}
    \mathcal{A}
    =
    N_{\rm pix}
    \sum_{i,j,k}\mathcal{K}_{ijk},
    \qquad
    \lambda^{\rm iso}_{ijk}
    =
    \phistar\,\frac{\mathcal{K}_{ijk}}{\mathcal{A}} .
    \label{eq:count-normalization}
\end{equation}
It follows immediately that
\begin{equation}
    \sum_{i,j,k,p}\lambda^{\rm iso}_{ijk}
    =
    N_{\rm pix}\sum_{i,j,k}\lambda^{\rm iso}_{ijk}
    =
    \phistar .
    \label{eq:phi-normalization}
\end{equation}
The parameter $\phistar$ is therefore the total expected source-count normalization across the full discretized sky and source grid.  The other source-population parameters have distinct roles: $\epsilon_0$ changes the mass-dependent scatter boost, $\MBHmax$ controls the high-mass cutoff, $\gamma$ controls the redshift evolution factor, and $\beta$ controls the frequency residence factor.  In the reduced analysis below, $\epsilon_0$ is fixed and $\gamma$ lies outside the posterior target set.

\subsection{Sky modulation and strain moments}
\label{subsec:lss}

The large-scale-structure field is represented as a shell-dependent angular contrast $\delta_j(\nhat_p)$ with zero monopole on each redshift shell.  It is generated from real spherical harmonics up to $\ell_{\rm max}=12$ and then mean-subtracted and rescaled so that the shell RMS follows the configured LSS amplitude times the growth factor \citep{Limber53,Kaiser92,Tegmark:2001xb,Tinker:2008ff,Tinker10}.  The important population-level operation is the modulation of the isotropic expected counts:
\begin{equation}
    \lambda_{ijkp}
    =
    \lambda^{\rm iso}_{ijk}\,
    \mathcal{M}_{jp}(b_{\rm BH}),
    \qquad
    \mathcal{M}_{jp}(b_{\rm BH})
    =
    \frac{\exp\left[b_{\rm BH}\delta_j(\nhat_p)\right]}
    {\left\langle \exp\left[b_{\rm BH}\delta_j(\nhat)\right]\right\rangle_{\nhat}} .
    \label{eq:lss-modulation}
\end{equation}
The denominator is evaluated as a sky-pixel average on each redshift shell.  Therefore
\begin{equation}
    \left\langle \mathcal{M}_{j}(b_{\rm BH}) \right\rangle_{\nhat}=1,
    \qquad
    \sum_p \lambda_{ijkp}
    =
    N_{\rm pix}\lambda^{\rm iso}_{ijk}.
    \label{eq:lss-mean-preserving}
\end{equation}
The parameter $b_{\rm BH}$ redistributes the shell-averaged source counts over the sky and primarily changes clustering-sensitive observables.  At $b_{\rm BH}=0$, $\mathcal{M}_{jp}=1$ and the count field is isotropic.  Increasing $b_{\rm BH}$ raises the angular contrast without changing the sky mean in a shell.  This angular response is considered separately from the three population coordinates inferred below.

\subsubsection*{Single-source strain and moment fields.}
\label{subsec:strain-moments}

Each cell is assigned a characteristic single-source strain amplitude using the detector-frame chirp mass \citep{Phinney:2001di,Sesana:2008mz}.  For a binary with component masses $m_1,m_2$, total mass $M=m_1+m_2$, and symmetric mass ratio $\eta=m_1m_2/M^2$, the usual chirp mass is $\Mc=\eta^{3/5}M$.  The fiducial convention uses the equal-mass value $\eta=1/4$, so that $\Mc_z=(1+z)\Mc$, $\Mc=q_{\Mc}M$, and $q_{\Mc}=2^{-6/5}=0.4352752816$.  This convention isolates the mass-loading response carried by the high-mass end of the population.  The adopted strain-amplitude convention is
\begin{equation}
    h_s(f,z,M)
    =
    \frac{
    2\left(G\Mc_z\right)^{5/3}
    \left(\pi f\right)^{2/3}
    }{
    c^4 d_L(z)
    } .
    \label{eq:single-source-strain}
\end{equation}
The expression is dimensionless.  The factor $(G\Mc_z)^{5/3}(\pi f)^{2/3}$ carries dimensions $m^5s^{-4}$; $c^4d_L$ carries the same dimensions.  We adopt this fixed half-amplitude convention relative to the common circular-binary amplitude $h_0=4(G\Mc_z)^{5/3}(\pi f)^{2/3}/(c^4d_L)$, so that $h_0=2h_s$.  The same normalization is used for all moment amplitudes, residuals, and power-law quantities.  Departures from the equal-mass convention rescale the strain as $h_s\propto\Mc_z^{5/3}\propto\eta M_z^{5/3}$ and $h_s^2\propto\eta^2M_z^{10/3}$.  A broad mass-ratio distribution would therefore change both the mean amplitude and the high-mass response and can be partly degenerate with source normalization and the mass cutoff \citep{Kelley:2016gse,NANOGrav:2023hfp,Liepold:2024woa}.  The present reference model fixes the mass-ratio convention.  With Poisson cell counts $N_{ijkp}\sim{\rm Pois}(\lambda_{ijkp})$, weighted population sums obey the compound-Poisson moment relations
\begin{align}
    \left\langle
    \sum_{j,k,p}N_{ijkp}h_s^2(f_i,z_j,M_k)
    \right\rangle
    &=
    \sum_{j,k,p}\lambda_{ijkp}h_s^2(f_i,z_j,M_k),
    \nonumber\\
    \left\langle
    \sum_{j,k,p}N_{ijkp}h_s^4(f_i,z_j,M_k)
    \right\rangle
    &=
    \sum_{j,k,p}\lambda_{ijkp}h_s^4(f_i,z_j,M_k).
    \label{eq:poisson-moments}
\end{align}
The corresponding cell moments used below are
\begin{align}
    \mu^{(2)}_{ijkp} &= \lambda_{ijkp}\,h_s^2(f_i,z_j,M_k), \nonumber\\
    \mu^{(4)}_{ijkp} &= \lambda_{ijkp}\,h_s^4(f_i,z_j,M_k).
    \label{eq:cell-moments}
\end{align}
The second moment is the binned strain-power weight used for the unresolved background map.  The fourth moment is a compound-Poisson participation moment of the power-weighted population; it controls finite-source and bright-tail dominance in the forward model \citep{Sato-Polito:2023spo,Sato-Polito:2024lew,Lamb:2024gbh,Allen:2022dzg,Allen:2024mtn}.  Summing over redshift and mass gives the per-frequency, per-pixel maps
\begin{align}
    H_{ip}^{(2)}
    &=
    \sum_{j,k}\mu^{(2)}_{ijkp},
    &
    H_{ip}^{(4)}
    &=
    \sum_{j,k}\mu^{(4)}_{ijkp}.
    \label{eq:moment-maps}
\end{align}
For an isotropic shell these reduce to the frequency moments
\begin{align}
    \overline{H}^{(2)}_i
    &=
    \sum_{j,k}\lambda^{\rm iso}_{ijk}h_s^2(f_i,z_j,M_k), \nonumber\\
    \overline{H}^{(4)}_i
    &=
    \sum_{j,k}\lambda^{\rm iso}_{ijk}h_s^4(f_i,z_j,M_k).
    \label{eq:iso-moments}
\end{align}
The per-pixel effective number of contributing sources and the dominant-cell fraction are then
\begin{align}
    \Neff^{\rm pix}(f_i)
    &=
    \frac{\left[\overline{H}^{(2)}_i\right]^2}
    {\overline{H}^{(4)}_i},
    \nonumber\\
    \eta_{\rm dom}(f_i)
    &=
    \frac{\max_{j,k}\left[\lambda^{\rm iso}_{ijk}h_s^2(f_i,z_j,M_k)\right]}
    {\overline{H}^{(2)}_i}.
    \label{eq:neff-dom}
\end{align}
These ratios are evaluated for positive second and fourth moments.  The scatter boost is restricted to $S_\epsilon(M)>0$, so all source-count weights are non-negative.  For an isotropic shell, the corresponding all-sky participation count is $\Neff^{\rm sky}(f_i)=N_{\rm pix}\Neff^{\rm pix}(f_i)$.  Increasing $\MBHmax$ extends the high-mass tail in Eq.~\eqref{eq:mass-kernel}; because $h_s^2\propto \Mc_z^{10/3}$ and $h_s^4\propto \Mc_z^{20/3}$, the mass tail affects bright-source dominance more strongly than it affects the total count.

The connection to the usual stochastic-background notation is through the frequency-binned strain power.  With the present equal-log-bin convention, the binned second moment is proportional to the characteristic-strain density times $\Delta\ln f_i$:
\begin{align}
    h_c^2(f_i)
    &\simeq
    \frac{1}{\Delta\ln f_i}
    \sum_{j,k,p}\lambda_{ijkp}h_s^2(f_i,z_j,M_k),
    \nonumber\\
    \Omega_{\rm GW}(f_i)
    &\propto
    f_i^2h_c^2(f_i).
    \label{eq:hc-omega-connection}
\end{align}
The constant $\Delta\ln f_i$ factor is absorbed in figures that use response ratios or local derivatives; those figures use a convention-matched amplitude scale for comparing response signs.
The conventional power-law description writes \citep{Phinney:2001di,NANOGrav:2023gor,EPTA:2023fyk,Reardon:2023gzh,Xu:2023wog}
\begin{equation}
    h_c(f)
    =
    A
    \left(\frac{f}{f_{\rm yr}}\right)^\alpha,
    \qquad
    P_{\rm gwb}(f)
    =
    \frac{h_c^2(f)}{12\pi^2 f^3},
    \label{eq:power-law-bridge}
\end{equation}
where $A$ is the characteristic-strain amplitude at $f_{\rm yr}=1\,{\rm yr}^{-1}$ and $\alpha$ is the strain spectral index.  We derive $h_c^2(f_i)$ from the binned second moment using the equal-log-bin convention in Eq.~\eqref{eq:hc-omega-connection}.  A line fitted to the four grid points between $1$ and $8\,{\rm nHz}$ is extrapolated to $f_{\rm yr}\simeq31.7\,{\rm nHz}$.  The resulting $A$ and $\alpha$ provide the comparison with standard PTA power-law quantities.  For any population parameter vector $\Theta$, the binned moments define
\begin{align}
    A^2(\Theta)
    &\simeq
    h_c^2(f_{\rm yr};\Theta),
    \nonumber\\
    \alpha(\Theta)
    &\simeq
    \frac{1}{2}\,
    \frac{\partial \ln H^{(2)}(f;\Theta)}
    {\partial \ln f}.
    \label{eq:A-alpha-map}
\end{align}
The second moment controls the mean strain-power level, and the fourth moment measures the contribution from rare bright cells.  These moments supply the information used below to compare the population coordinates with the conventional amplitude, spectral index, and source-participation quantities.

\subsection{Foreground, background, and PTA response}
\label{subsec:foreground-background}

The simulation distinguishes individually sampled bright foreground cells from the unresolved background using the same moments \citep{Sato-Polito:2024lew,Lamb:2024gbh,Agazie:2024jbf}.  For a background mask $B_{ijkp}\in\{0,1\}$ and foreground mask $F_{ijkp}=1-B_{ijkp}$, the background strain-power and fourth-moment maps are
\begin{align}
    H^{(2),{\rm bg}}_{ip}
    &=
    \sum_{j,k}B_{ijkp}\lambda_{ijkp}h_s^2(f_i,z_j,M_k),
    \nonumber\\
    H^{(4),{\rm bg}}_{ip}
    &=
    \sum_{j,k}B_{ijkp}\lambda_{ijkp}h_s^4(f_i,z_j,M_k).
    \label{eq:bg-moments}
\end{align}
The corresponding local background effective count and dominant fraction are
\begin{align}
    \Neff^{\rm bg}(f_i,\nhat_p)
    &=
    \frac{\left[H^{(2),{\rm bg}}_{ip}\right]^2}
    {H^{(4),{\rm bg}}_{ip}},
    \nonumber\\
    \eta_{\rm dom}^{\rm bg}(f_i,\nhat_p)
    &=
    \frac{\max_{j,k}B_{ijkp}\lambda_{ijkp}h_s^2(f_i,z_j,M_k)}
    {H^{(2),{\rm bg}}_{ip}}.
    \label{eq:bg-neff-dom}
\end{align}
Cells are moved into the foreground when a single cell would dominate the remaining local background or when the local effective count is too small, subject to a minimum retained fraction of the background strain power.  The adopted split uses a source-to-background power threshold $8.0\times10^{-3}$, an effective-count threshold $\Neff=2.0$, and a minimum retained background-power fraction $0.25$.  This moment-based definition separates individually important sources from the unresolved stochastic map.

\subsubsection*{PTA response and pair cross-power.}
\label{subsec:pta-response}

The sky maps are projected into pulsar-pair observables through the tensor antenna pattern \citep{Hellings:1983fr,Mingarelli:2013dsa,Taylor:2013esa,Cornish:2013aba,Gair:2015hra}.  For a source direction $\nhat$ and propagation direction $\khat=-\nhat$, choose transverse basis vectors $\uhat,\vhat\perp\khat$.  For pulsar $a$ with unit direction $\phat_a$, the response is
\begin{align}
    F_a^+(\nhat)
    &=
    \frac{1}{2}
    \frac{
    (\phat_a\cdot\uhat)^2-(\phat_a\cdot\vhat)^2
    }{
    1+\phat_a\cdot\khat
    },
    \nonumber\\
    F_a^\times(\nhat)
    &=
    \frac{
    (\phat_a\cdot\uhat)(\phat_a\cdot\vhat)
    }{
    1+\phat_a\cdot\khat
    }.
    \label{eq:antenna-pattern}
\end{align}
The pixel response kernel for a pulsar pair $(a,b)$ is
\begin{equation}
    \Gamma_{ab,p}
    =
    F_a^+(\nhat_p)F_b^+(\nhat_p)
    +
    F_a^\times(\nhat_p)F_b^\times(\nhat_p).
    \label{eq:pair-kernel}
\end{equation}
Polarization-averaging and overall response-normalization factors are absorbed into this convention.  With this normalization, an isotropic sky recovers the Hellings--Downs angular dependence up to the same convention factors used in the timing-residual summaries; the analysis below uses relative response information throughout.
The unresolved background pair cross-power at frequency $f_i$ can be written in two equivalent conventions.  If the map is represented as a pixel-integrated moment $H^{(2),{\rm bg}}_{ip}$, the pixel area is already included.  If it is represented as an intensity per steradian, $I^{(2),{\rm bg}}_{ip}=H^{(2),{\rm bg}}_{ip}/\Delta\Omega$, the area appears explicitly:
\begin{equation}
    C_{ab}(f_i)
    =
    \sum_p
    I^{(2),{\rm bg}}_{ip}\,
    \Gamma_{ab,p}\,
    \Delta\Omega
    =
    \sum_p
    H^{(2),{\rm bg}}_{ip}\,
    \Gamma_{ab,p}.
    \label{eq:pair-cross-power}
\end{equation}
The normalization choice is kept consistent throughout the analysis; Appendix~\ref{app:angular-frontier} compares the isotropic convention against the Hellings--Downs template.  The isotropic limit of Eq.~\eqref{eq:pair-cross-power} recovers the familiar Hellings--Downs angular dependence after the sky average.  For pulsar angular separation $\zeta$,
\begin{equation}
    \Gamma_{\rm HD}(\zeta)
    =
    \frac{1}{2}
    +
    \frac{3}{2}x\ln x
    -
    \frac{x}{4},
    \qquad
    x=\frac{1-\cos\zeta}{2},
    \label{eq:hellings-downs}
\end{equation}
with the autocorrelation diagonal set separately to unity.  This convention fixes the pair-response normalization used in the residual-level summaries.

\subsection{Residual-level observables and raw parameter roles}
\label{subsec:residuals}

At the timing-residual level, a foreground source with angular frequency $\omega=2\pi f$ contributes an Earth term and, when included, a delayed pulsar term.  For inclination cosine $\iota_c=\cos\iota$, polarization angle $\psi$, initial orbital angle $\varphi_0$, and rotated antenna responses $\widetilde{F}_a^+$ and $\widetilde{F}_a^\times$, the single-source residual has the form
\begin{align}
    r_{a,E}(t)
    &=
    \frac{h_s}{\omega}
    \left[
    \widetilde{F}_a^+
    \frac{1+\iota_c^2}{2}
    \sin(\omega t+\varphi_0)
    \right.
    \nonumber\\
    &\quad\left.
    +
    \widetilde{F}_a^\times
    \iota_c
    \cos(\omega t+\varphi_0)
    \right],
    \nonumber\\
    r_{a,P}(t)
    &=
    -\frac{h_s}{\omega}
    \left[
    \widetilde{F}_a^+
    \frac{1+\iota_c^2}{2}
    \sin(\omega[t-\tau_a]+\varphi_0)
    \right.
    \nonumber\\
    &\quad\left.
    +
    \widetilde{F}_a^\times
    \iota_c
    \cos(\omega[t-\tau_a]+\varphi_0)
    \right],
    \label{eq:foreground-residual}
\end{align}
with pulsar-term delay $\tau_a=L_a(1+\khat\cdot\phat_a)/c$.  Here $L_a$ is the pulsar distance; if distances are stored in light-travel-time units the factor of $c$ is absorbed into $L_a$.
The scale of Eq.~\eqref{eq:foreground-residual} is $|r|\sim h_s/(2\pi f)$, which is why nanohertz sources can produce measurable timing residuals despite very small strain amplitudes: the timing residual integrates the redshift signal over a long wave period.  Lower observed frequency gives a larger residual response at fixed strain amplitude.

For the unresolved background, the residual model samples shared complex sky coefficients for plus and cross polarizations and projects the same realization into each pulsar.  The pulsar contribution includes the propagation delay to each pulsar.  The construction retains the common Earth term and the delayed pulsar terms and focuses the analysis on the population moments and PTA response.

\subsubsection*{Physical meaning of the raw parameters.}
\label{subsec:raw-parameter-meaning}

The raw population parameters enter the model in different physical places:
\begin{itemize}
    \item $\phistar$ fixes the global expected source count through Eq.~\eqref{eq:phi-normalization}.
    \item $\epsilon_0$ changes the mass-dependent scatter boost in Eq.~\eqref{eq:scatter-boost}.
    \item $\MBHmax$ controls the high-mass cutoff in Eq.~\eqref{eq:mass-kernel}.
    \item $\beta$ controls the frequency residence factor in Eq.~\eqref{eq:residence-time}.
    \item $\gamma$ controls the redshift evolution in Eq.~\eqref{eq:redshift-kernel}.
    \item $b_{\rm BH}$ controls the angular LSS modulation in Eq.~\eqref{eq:lss-modulation}.
\end{itemize}
The simulated observables respond to correlated combinations of these raw parameters.  The next section defines the three population coordinates used to represent those combinations.

\section{Population Coordinates and Posterior Inference}
\label{sec:methodology}
\label{sec:population-coordinates}

\subsection{Population-coordinate target}
\label{subsec:inference-object}

The raw population coordinates considered in this analysis are
\begin{equation}
    \Theta_{\rm raw}
    =
    \left(
    \beta,\ \phistar,\ \log_{10}\frac{\MBHmax}{10^{10}\,\Msun},\ b_{\rm BH}
    \right).
    \label{eq:raw-coordinates}
\end{equation}
Here $\beta$ controls the frequency residence factor, $\phistar$ sets the expected source count, $\MBHmax$ controls the high-mass cutoff, and $b_{\rm BH}$ modulates the sky distribution.  The posterior target uses three combinations that follow the principal spectral, normalization, and high-mass responses:
\begin{equation}
    \Theta_{\rm eff}
    =
    \left(
    \beta,\ \phieff,\ \meff
    \right).
    \label{eq:population-target}
\end{equation}
The angular coordinate $b_{\rm BH}$ is retained in the forward model and studied through its response in Sec.~\ref{app:angular-frontier}.  The posterior target in Eq.~\eqref{eq:population-target} contains the three sky-averaged population coordinates.

\begin{table}
    \centering
    \caption{Population coordinates used as posterior targets.}
    \label{tab:population-targets}
    \resizebox{\linewidth}{!}{%
        \begin{tabular}{llll}
        \hline
Target & Definition & Physical interpretation & Principal sensitivity \\
        \hline
        $\beta$ & raw $\beta$ & residence time and spectral shape & frequency-resolved observables \\
        $\phieff$ & $0.707\,z(\phistar)-0.707\,z(\beta)$ & source normalization at fixed residence response & foreground/background moments \\
        $\meff$ & $0.181\,z(\phistar)+0.983\,z(\ellM)$ & high-mass source contribution & finite-source moments \\
        \hline
    \end{tabular}
    }
\end{table}

\subsection{Simulation-based posterior estimation}
\label{subsec:posterior-estimation}

The observable vector $x_{\rm PTA}$ contains frequency-dependent strain summaries, pulsar-pair correlations, foreground and background moments, finite-source participation, and angular responses after the observing and residual treatment.  We approximate $p(\Theta_{\rm eff}\mid x_{\rm PTA})$ with neural posterior estimation using a normalizing flow \citep{Cranmer:2020sbi,TejeroCantero:2020sbi,Papamakarios:2019snl,Greenberg:2019apt,Durkan:2019nsf}.  The reference ensemble contains $N=16384$ simulations.  A separate $N=256$ evaluation ensemble is used for coordinate selection, recovery, coverage, posterior geometry, and the matched comparisons reported below.

For each evaluation realization we report posterior means $\bar{\theta}_{n,a}$, marginal widths $\sigma_{n,a}$, posterior-centred residuals $z^{\rm post}_{n,a}=(\theta^{\rm true}_{n,a}-\bar{\theta}_{n,a})/\sigma_{n,a}$, and empirical central coverage.  The central 90 per cent coverage is
\begin{equation}
    C_{90,a}
    =
    \frac{1}{N_{\rm eval}}
    \sum_{n=1}^{N_{\rm eval}}
    \mathbb{I}
    \left[
    q_{0.05,n,a}
    \le
    \theta^{\rm true}_{n,a}
    \le
    q_{0.95,n,a}
    \right],
    \label{eq:coverage90}
\end{equation}
where $q_{0.05,n,a}$ and $q_{0.95,n,a}$ are posterior quantiles.  Point recovery is summarized by the Pearson correlation between $\bar{\theta}_{n,a}$ and $\theta^{\rm true}_{n,a}$ and by the normalized mean absolute error
\begin{equation}
    {\rm NMAE}_a
    =
    \frac{
    N_{\rm eval}^{-1}\sum_n
    \left|\bar{\theta}_{n,a}-\theta^{\rm true}_{n,a}\right|
    }{
    {\rm std}\left(\theta^{\rm true}_{a}\right)
    }.
    \label{eq:nmae}
\end{equation}

\subsection{Comparison coordinates}
\label{subsec:traditional-coordinate-comparison}

For comparison with the usual PTA common-process description, we also construct $\Theta_{\rm trad}=(\log_{10}A,\gamma_{\rm cp},\log_{10}\Neff)$.  The amplitude and spectral index come from a four-bin power-law fit to the binned second strain moment, and $\Neff$ is formed from the foreground second and fourth moments.  These quantities connect the population calculation to common-process and free-spectrum analyses \citep{NANOGrav:2023gor,EPTA:2023fyk,Reardon:2023gzh,Xu:2023wog,InternationalPulsarTimingArray:2023mzf}.  Section~\ref{subsec:fig8-forecast} and Appendix~\ref{app:matched-coordinate-comparisons} compare the two three-coordinate descriptions using matched simulations.

\subsection{Standardized coordinate convention}
\label{subsec:standardized-coordinates}

The two derived coordinates are defined in standardized raw-coordinate planes.  For any scalar raw coordinate $x$,
\begin{equation}
    z(x)
    =
    \frac{x-\mu_x}{s_x},
    \label{eq:z-standardization}
\end{equation}
where $\mu_x$ and $s_x$ are measured from the reference simulation ensemble and then held fixed.

Table~\ref{tab:standardization-constants} gives the constants used throughout the analysis.

\begin{table}
    \centering
    \caption{Standardization constants used in the population-coordinate definitions.}
    \label{tab:standardization-constants}
    \begin{tabular}{lcc}
        \hline
        Raw coordinate & Mean $\mu_x$ & Scale $s_x$ \\
        \hline
        $\beta$ & $2.457002$ & $0.373012$ \\
        $\phistar$ & $128.603237$ & $41.587761$ \\
        $\ell_M$ & $0.012882$ & $0.135960$ \\
        \hline
    \end{tabular}
\end{table}

\subsection{Normalization coordinate \texorpdfstring{$\phieff$}{phi-eff}}
\label{subsec:phi-eff-method}

The normalization coordinate is selected in the standardized $(\phistar,\beta)$ plane from the family
\begin{align}
    u_\phi(\alpha)
    &=
    w_\phi(\alpha)\,z(\phistar)
    +
    w_\beta(\alpha)\,z(\beta),
    \nonumber\\
    w_\phi=\cos\alpha,\quad
    &w_\beta=\sin\alpha,
    \label{eq:phi-search-family}
\end{align}
with $w_\phi>0$, $w_\beta<0$, and $|w_\phi|\ge |w_\beta|$.  These conditions select a normalization-dominated combination that subtracts the covariance with $\beta$.  Point recovery, posterior width, and coverage in the evaluation ensemble select $\alpha=-45^\circ$, giving
\begin{equation}
    \boxed{
    \phieff
    =
    \frac{1}{\sqrt{2}}\,z(\phistar)
    -
    \frac{1}{\sqrt{2}}\,z(\beta)
    }.
    \label{eq:phi-eff-definition}
\end{equation}
The resulting point-recovery correlation is $r=0.926$ and the NMAE is $0.150$.  The central 90 per cent coverage is $0.871$, below the nominal value by $0.029$.  The broader posterior for $\phieff$ reflects the remaining covariance between residence time and source normalization.  The response plots vary the corresponding raw parameters along the vector defined by Eq.~\eqref{eq:phi-eff-definition}.

\subsection{High-mass coordinate \texorpdfstring{$\meff$}{m-eff}}
\label{subsec:m-eff-method}

Define $\ellM=\log_{10}(\MBHmax/10^{10}\,\Msun)$.  The high-mass coordinate is selected in the standardized $(\phistar,\ellM)$ plane from
\begin{equation}
    u_M(a,b)
    =
    a\,z(\phistar)+b\,z(\ellM),
    \qquad
    a^2+b^2=1.
    \label{eq:m-search-family}
\end{equation}
Minimizing the projected NMAE of the posterior mean in the evaluation ensemble gives the rounded coefficients $a=0.181$ and $b=0.983$:
\begin{equation}
    \boxed{
    \meff
    =
    0.181\,z(\phistar)
    +
    0.983\,z\!\left[
    \log_{10}\left(\frac{\MBHmax}{10^{10}\,\Msun}\right)
    \right]
    }.
    \label{eq:m-eff-definition}
\end{equation}
In the coordinate-selection calculation, $\meff$ has ${\rm NMAE}=0.0879$, Pearson $r=0.9923$, and median-split accuracy $0.9805$.  The corresponding NMAE values are $0.2244$ for $\phistar$ and $0.0960$ for $\ellM$.  In the posterior recovery calculation used for Fig.~\ref{fig:recovered-vs-truth}, $\meff$ has $r=0.884$ and $C_{90}=0.898\pm0.019$.  Its coefficient on $\ellM$ identifies it primarily with the high-mass cutoff, with a smaller contribution from source normalization.

\subsection{Posterior geometry and local separation}
\label{subsec:information-geometry}

For a raw parameter vector $\Theta_{\rm raw}$ and posterior covariance $\Sigma_{\Theta}$, the local information matrix is approximated by $\mathcal{I}_{\Theta}\simeq\Sigma_{\Theta}^{-1}$.  In a standardized two-parameter subspace, an information-aligned vector satisfies $\mathcal{I}_{\rm sub}\bm{w}_a=\lambda_a\bm{w}_a$.  Across 32 evaluation posteriors, $\meff$ has median alignment $0.997$ with the high-mass precision axis, with a 16--84 per cent range of $0.995$--$0.998$.  The corresponding alignment for $\phieff$ is $0.729$, with a range of $0.713$--$0.744$.  The geometry is therefore more strongly aligned with the high-mass coordinate than with the normalization coordinate.

We also compare posterior widths with the local spacing of simulated population values.  For a two-coordinate plane $(a,b)$, let $s_a={\rm std}(\theta_a^{\rm true})$ and $s_b={\rm std}(\theta_b^{\rm true})$, and place each evaluation realization at $\bm y_n=(\theta_{n,a}^{\rm true}/s_a,\theta_{n,b}^{\rm true}/s_b)$.  Let $d_n$ be the median distance from $\bm y_n$ to its six nearest neighbours.  The local posterior radius and posterior-mean offset are
\begin{align}
    R_n
    &=
    \left[
    \left(\frac{\sigma_{n,a}}{s_a}\right)^2+
    \left(\frac{\sigma_{n,b}}{s_b}\right)^2
    \right]^{1/2},
    \nonumber\\
    B_n
    &=
    \left[
    \left(\frac{\bar{\theta}_{n,a}-\theta_{n,a}^{\rm true}}{s_a}\right)^2+
    \left(\frac{\bar{\theta}_{n,b}-\theta_{n,b}^{\rm true}}{s_b}\right)^2
    \right]^{1/2}.
    \label{eq:radius-bias}
\end{align}
The separation score is $D_n=d_n/R_n$.  We count an evaluation point as locally separated when $D_n\ge1.5$ and $B_n\le1.0$.  Its interpretation is restricted to a relative comparison among the three coordinate planes.

\section{Results}
\label{sec:validation-results}
\label{sec:coordinate-results}

We first examine the forward response of the three coordinates, then their geometry in the raw-parameter planes and their recovery in the evaluation simulations.

\subsection{Observable responses}
\label{subsec:fig2-response}

\begin{figure*}
    \centering
    \includegraphics[width=\textwidth]{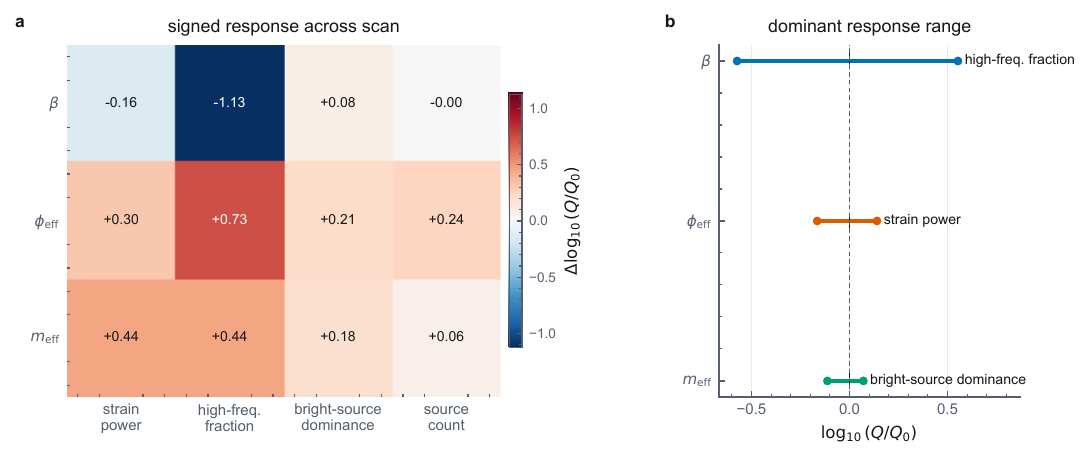}
\caption{Responses of the population coordinates.  The matrix shows signed changes in $\log_{10}(Q/Q_0)$ along each one-dimensional scan.  The right panel gives the high-frequency fraction for $\beta$, integrated strain power for $\phieff$, and bright-source dominance for $\meff$.}
    \label{fig:parameter-response}
\end{figure*}

For a scalar observable $Q$ measured along a one-dimensional parameter scan, Fig.~\ref{fig:parameter-response} shows $\mathcal{R}_Q(\Delta\theta)=Q(\Delta\theta)/Q(0)$.  Each coordinate affects several observables.  Their largest changes occur in different quantities:
\begin{itemize}
    \item $\beta$ changes the spectral response through the residence factor in Eq.~\eqref{eq:residence-time}; the high-frequency fraction spans $0.267$--$3.573$ times its reference value.
    \item $\phieff$ changes the source-normalization response; the integrated strain power spans $0.684$--$1.380$ times its reference value.
    \item $\meff$ changes the contribution from massive bright sources; the bright-source dominance spans $0.772$--$1.177$ times its reference value.
\end{itemize}

\subsection{Coordinate geometry}
\label{subsec:fig3-geometry}

\begin{figure*}
    \centering
    \includegraphics[width=\textwidth]{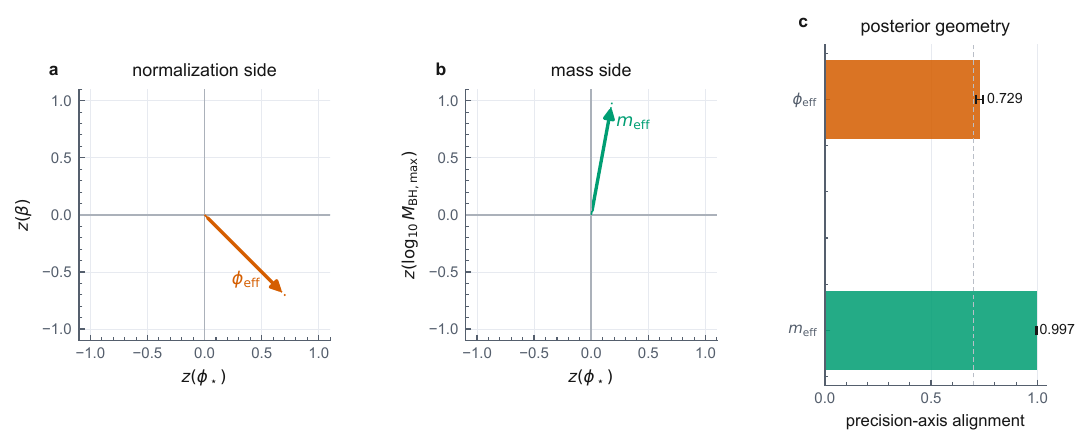}
\caption{Population coordinates in standardized raw-parameter planes.  The left panel shows $\phieff$ in the $\beta$--$\phistar$ plane, and the right panel shows $\meff$ in the $\phistar$--$\ellM$ plane.  The side strip gives their alignment with the local posterior geometry.}
    \label{fig:coordinate-geometry}
\end{figure*}

\begin{figure*}
    \centering
    \includegraphics[width=\textwidth]{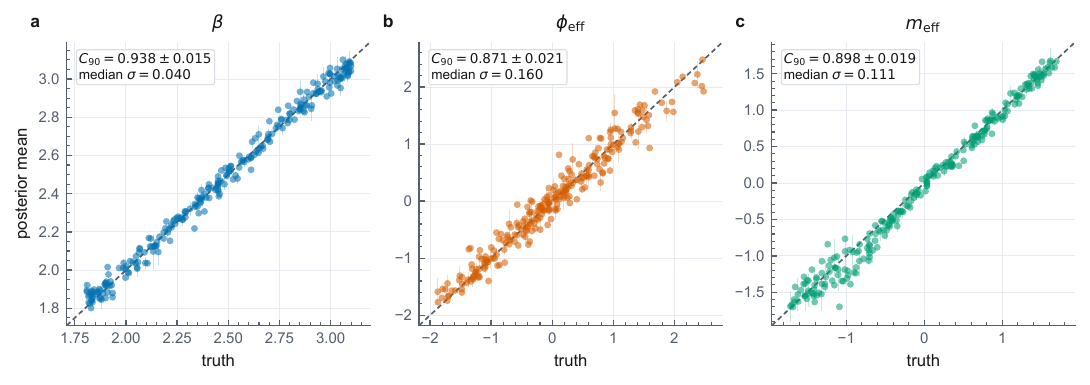}
\caption{Recovery of the three population coordinates in the evaluation simulations.  Each panel compares the posterior mean with the simulated value and gives the empirical central 90 per cent coverage and median marginal posterior width.}
    \label{fig:recovered-vs-truth}
\end{figure*}

Figure~\ref{fig:coordinate-geometry} places $\phieff$ and $\meff$ in their standardized raw-parameter planes.  The selected normalization vector has point-recovery correlation $r\simeq0.926$.  The high-mass vector gives projected NMAE $0.0872$, compared with $0.2244$ for $\phistar$ and $0.0960$ for $\ell_M$.  Its dominant coefficient on $\ell_M$ and its alignment with the posterior precision axis identify $\meff$ with the high-mass response.

\subsection{Recovery in simulated observations}
\label{subsec:fig4-recovery}

The posterior-mean correlations are $r_\beta=0.928$, $r_{\phieff}=0.926$, and $r_{\meff}=0.884$.  The empirical central 90 per cent coverages are $C_{90,\beta}=0.938\pm0.015$, $C_{90,\phieff}=0.871\pm0.021$, and $C_{90,\meff}=0.898\pm0.019$, where the uncertainties are binomial standard errors for $N=256$ evaluation realizations.  The lower coverage and broader posterior for $\phieff$ are consistent with the remaining covariance between residence time and source normalization.

\section{Physical Interpretation and Discussion}
\label{sec:physical-picture}
\label{sec:implications-outlook}

\subsection{Relative local separation and observable sensitivity}
\label{subsec:fig6-distinguishability}

Table~\ref{tab:fig6-plane-summary} compares the local spacing of simulated values with the posterior radii defined in Sec.~\ref{subsec:information-geometry}.  The $\beta-\meff$ plane has the largest median score, $D=1.196$, and a locally separated fraction of $0.301\pm0.029$.  The corresponding fractions are $0.090\pm0.018$ for $\beta-\phieff$ and $0.273\pm0.028$ for $\phieff-\meff$.  Most evaluation realizations therefore overlap their neighbours at the adopted sensitivity.  Among the three planes, changes in residence time and high-mass contribution give the largest relative separation.

\begin{table}
    \centering
    \caption{Local posterior-radius separation for the three population-coordinate planes.}
    \label{tab:fig6-plane-summary}
    \resizebox{\columnwidth}{!}{%
    \begin{tabular}{lcc}
        \hline
        Plane & median $D$ & $f_{\rm sep}$ \\
        \hline
        $\beta-\phieff$ & $0.877$ & $0.090\pm0.018$ \\
        $\beta-\meff$ & $1.196$ & $0.301\pm0.029$ \\
        $\phieff-\meff$ & $1.017$ & $0.273\pm0.028$ \\
        \hline
    \end{tabular}
    }
\end{table}

\begin{figure*}
    \centering
    \includegraphics[width=0.72\textwidth]{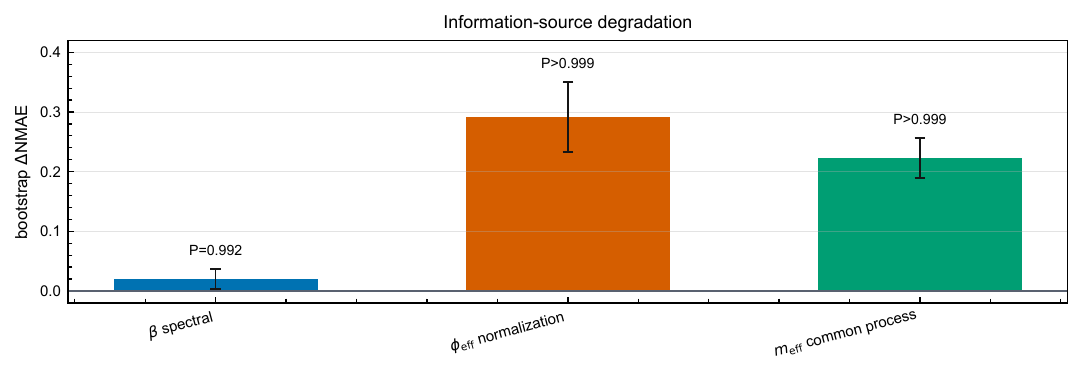}
\caption{Sensitivity to selected observable summaries.  The bars show paired-bootstrap changes in target-level NMAE when frequency-resolved observables or the foreground/background split are removed, or when the input is compressed to common-process quantities.}
    \label{fig:summary-family-attribution}
\end{figure*}

Figure~\ref{fig:summary-family-attribution} quantifies sensitivity to selected families of observables.  Removing explicit frequency information increases the $\beta$ NMAE by $0.019$ with paired-bootstrap probability $P=0.992$.  Removing the foreground/background split increases the $\phieff$ NMAE by $0.291$, and compression to common-process quantities increases the $\meff$ NMAE by $0.223$; both changes have $P>0.999$.  These tests identify influential summary families, although each family contains several correlated observables.

\subsection{High-mass finite-source physics}
\label{subsec:meff-physical-attribution}

\begin{figure*}
    \centering
    \includegraphics[width=\textwidth]{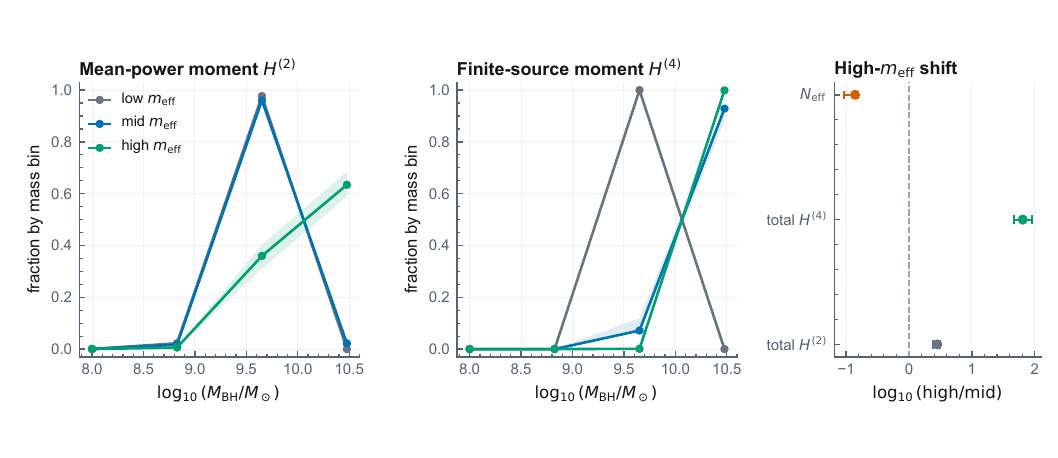}
\caption{Mass contributions associated with $\meff$.  The left and middle panels compare the median fractional contributions to the second and fourth strain moments for the three $\meff$ tertiles.  The right panel gives high-to-middle ratios for total $H^{(2)}$, total $H^{(4)}$, and $N_{\rm eff}$.}
    \label{fig:meff-mass-bin-attribution}
\end{figure*}

Figure~\ref{fig:meff-mass-bin-attribution} shows how the high-mass tertile in $\meff$ shifts strain power toward the largest mass bin.  Its median contribution to $H^{(2)}$ is $0.634$, compared with $0.022$ in the middle tertile.  The high-to-middle median ratios are $2.82$ for total $H^{(2)}$ and $64.4$ for total $H^{(4)}$, and the participation proxy $N_{\rm eff}=(\sum H^{(2)})^2/\sum H^{(4)}$ decreases to a ratio of $0.139$.  The stronger fourth-moment response shows that $\meff$ traces rare massive sources more directly than the mean strain power alone.

\subsection{Relation to power-law and common-process coordinates}
\label{subsec:fig8-forecast}

\subsubsection{Residence time and spectral index}
\label{subsec:analytic-bridge}
At fixed redshift and binary mass, Eq.~\eqref{eq:single-source-strain} gives $h_s^2\propto f^{4/3}$, and the source occupancy per logarithmic interval scales as $dN/d\ln f\propto f^{-\beta}$.  The binned second moment therefore scales as $H^{(2)}(f)\propto f^{4/3-\beta}$.  This gives $h_c(f)\propto f^{2/3-\beta/2}$ and
\begin{equation}
    \alpha \simeq \frac{2}{3}-\frac{\beta}{2}.
    \label{eq:analytic-alpha-bridge}
\end{equation}
With the PTA timing residual convention $\gamma_{\rm cp}=3-2\alpha$, this gives
\begin{equation}
    \gamma_{\rm cp}\simeq \frac{5}{3}+\beta .
    \label{eq:analytic-gamma-bridge}
\end{equation}
The GW driven value $\beta_{\rm GW}=8/3$ maps to $\gamma_{\rm cp}=13/3$.

\begin{figure*}
    \centering
    \includegraphics[width=\textwidth]{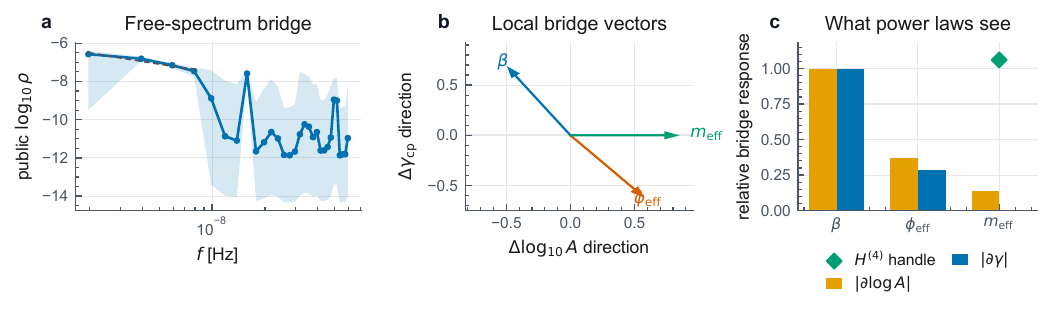}
\caption{Comparison with the NANOGrav 15-year free spectrum.  The left panel shows the public free-spectrum KDE in its published $\log_{10}\rho$ convention and a low-frequency linear fit.  The middle and right panels show the responses of the population coordinates in the amplitude--spectral-index plane.}
    \label{fig:public-pta-bridge}
\end{figure*}

The full binned calculation gives $\alpha=-0.667$ and $\gamma_{\rm cp}=4.333$ at the reference point.  The coordinate $\beta$ changes the spectral index most strongly, $\phieff$ primarily changes the amplitude, and $\meff$ has nearly zero spectral-index response.  Figure~\ref{fig:public-pta-bridge} provides an observational comparison of these amplitude and spectral-index responses with the public NANOGrav 15-year free spectrum.

\begin{figure*}
    \centering
    \includegraphics[width=0.92\textwidth]{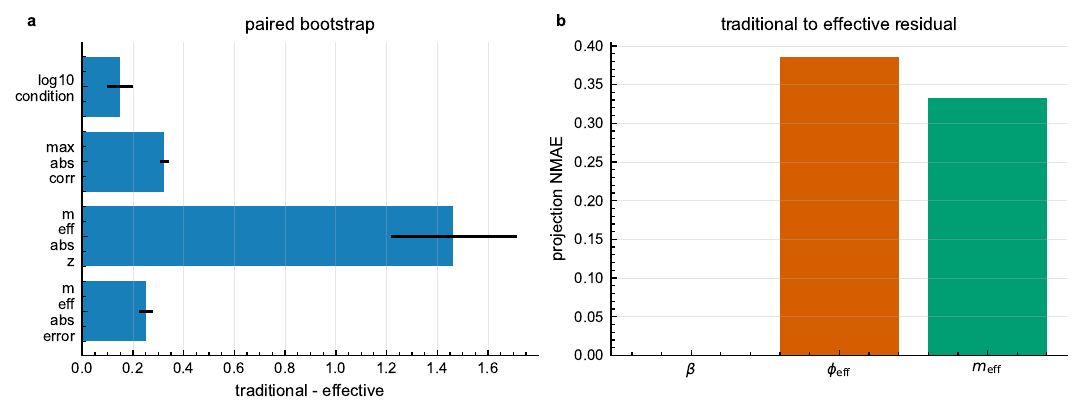}
\caption{Matched comparison with common-process coordinates.  The panels compare direct recovery of $(\beta,\phieff,\meff)$ with the projection of the $(A,\gamma_{\rm cp},N_{\rm eff})$ posterior into the same coordinates.  Error bars show paired-bootstrap intervals.}
    \label{fig:common-process-comparison}
\end{figure*}

Figure~\ref{fig:common-process-comparison} gives the matched comparison with the traditional posterior, which has $C_{90}=0.977$ for $\log_{10}A$, $0.996$ for $\gamma_{\rm cp}$, and $0.922$ for $\log_{10}\Neff$.  Projection of its samples into $(\beta,\phieff,\meff)$ gives $\meff$ coverage $0.480$, compared with $0.875$ for direct inference of the population coordinates.  The direct calculation reduces the $\meff$ absolute error by $0.251$, with a paired-bootstrap 95 per cent interval of $[0.223,0.279]$.  The amplitude and spectral index accurately summarize the mean common process, and the population coordinates retain additional information from the mass-dependent finite-source observables.

\subsection{Sensitivity to population and observing assumptions}
\label{subsec:robustness-scope}

Appendix~\ref{app:sensitivity} collects the supplementary calculations.  The coordinate conventions and selection stability are described in Appendix~\ref{app:coordinate-scope}.  Appendix~\ref{app:model-observing-sensitivity} examines grid resolution, the foreground/background threshold, representative PTA observing configurations, and population variants based on published models.  The population variants change the host relation, high-mass tail, mass-ratio mapping, eccentricity, environmental response, and redshift delay \citep{Kormendy:2013dxa,McConnell:2011mu,McConnell:2012hz,Li24,Yuan:2023ezi,Kelley:2016gse,NANOGrav:2023hfp,IzquierdoVillalba:2022gcs,Sykes:2022gwb,Bonetti:2018sec}.  The observing configurations represent NANOGrav, EPTA, PPTA, IPTA, and an SKA forecast \citep{NANOGrav:2023timing,NANOGrav:2023noise,EPTA:2023data,PPTA:2023dr3,IPTA:2019dr2,Janssen:2015sqa}.  These calculations test the signs and relative magnitudes of the forward responses; they do not constitute population inference from those data sets.  Appendix~\ref{app:width-rescaling} records the illustrative observing-span rescaling used to indicate how posterior widths affect local separation.

\clearpage
\onecolumn
\begin{multicols}{2}
\subsection{Conclusions}
\label{sec:conclusion}

We have developed a population inference for the nanohertz SMBHB signal that follows source abundance, binary residence time, the high-mass population, finite-source strain moments, and the PTA response in a common forward calculation.  The observable summaries identify three population coordinates associated with distinct physical responses: $\beta$ for residence time and spectral shape, $\phieff$ for source normalization after accounting for its covariance with $\beta$, and $\meff$ for the high-mass contribution.  This coordinate set expresses the combinations measured by the simulated PTA observables and connects them to the astrophysical ingredients of the source population.

Across the $N=256$ evaluation ensemble, the posterior-mean correlations with the simulated values are $0.928$ for $\beta$, $0.926$ for $\phieff$, and $0.884$ for $\meff$.  Their empirical central 90 per cent coverages are $0.938\pm0.015$, $0.871\pm0.021$, and $0.898\pm0.019$, respectively.  The coordinate construction gives ${\rm NMAE}=0.150$ for $\phieff$ and a projected ${\rm NMAE}=0.0879$ for $\meff$.  The broader $\phieff$ posterior and its lower coverage arise from the remaining covariance between residence time and population normalization.  Posterior precision and the spacing of nearby population realizations together determine the practical resolution of the coordinates, and most neighbouring simulations overlap at the sensitivity represented by this ensemble.

The forward responses give each coordinate a direct physical interpretation.  The exponent $\beta$ changes the residence-time weighting across frequency and maps to the common-process spectral index through $\gamma_{\rm cp}\simeq5/3+\beta$.  The coordinate $\phieff$ isolates the normalization response after subtracting its leading covariance with $\beta$.  The dominant coefficient of $\meff$ follows the high-mass cutoff, and its median alignment with the high-mass posterior precision axis is $0.997$.  High-$\meff$ populations increase the second strain moment and produce a much larger change in the fourth moment, linking this coordinate to rare, massive binaries.  The $\beta-\meff$ plane consequently has the largest relative local separation, with median $D=1.196$ and locally separated fraction $0.301\pm0.029$.

The comparison with conventional PTA coordinates clarifies how the available observables divide the population information.  Amplitude and spectral index summarize the mean common process efficiently.  Frequency-resolved summaries contribute most directly to $\beta$, the foreground/background moments affect the normalization coordinate, and mass-dependent finite-source statistics improve recovery of $\meff$.  In the matched common-process calculation, direct population-coordinate inference gives $\meff$ coverage of $0.875$, compared with $0.480$ after projecting the common-process posterior, and reduces the absolute error by $0.251$.  The fixed three-dimensional comparison also lowers the median maximum posterior correlation from $0.455$ to $0.383$ and the median log covariance condition from $5.68$ to $1.21$.  Tests with representative PTA configurations retain the response signs of $\beta$ and $\meff$, supporting their physical interpretation across the observing assumptions considered here.

The numerical coefficients of $\phieff$ and $\meff$ are tied to the phenomenological population family, the standardization convention, and the selected PTA summaries.  Substantial local posterior overlap sets the current resolution of population differences within this model.  Higher-fidelity treatments of galaxy mergers, black-hole--host relations, eccentricity, environmental hardening, and redshift evolution can determine how the measured combinations change as the source model is broadened.  Application to PTA data will also benefit from longer observing spans, improved noise characterization, and joint use of the free spectrum, continuous-wave candidates, anisotropy, and higher finite-source statistics.  These developments can extend the same response-based construction from the simulated population coordinates studied here to increasingly detailed measurements of SMBHB assembly and evolution.

\section*{Data Availability}
The code and paper-related materials are publicly available at \url{https://github.com/l46430640-del/PTA_HGNN_SBI_repository}.  The public PTA data products and catalogues used for comparison are available from the sources cited in the text.
\end{multicols}

\clearpage
\begingroup
\setlength{\columnsep}{18pt}
\makeatletter
\let\mnras@thebibliography\thebibliography
\renewcommand{\thebibliography}[1]{%
  \mnras@thebibliography{#1}%
  \fontsize{7}{7.8}\selectfont
  \setlength{\parskip}{0pt}%
  \setlength{\itemsep}{0pt}%
  \setlength{\parsep}{0pt}%
}
\makeatother
\begin{multicols}{2}
\bibliographystyle{mnras}
\bibliography{references}
\end{multicols}
\endgroup

\twocolumn
\appendix
\section{Sensitivity to Model and Observing Assumptions}
\label{app:sensitivity}

This appendix gives the coordinate conventions, posterior calibration, local-separation results, matched three-dimensional comparisons, angular response, and illustrative observing-span rescaling.

\subsection{Coordinate definitions}
\label{app:coordinate-scope}

The coordinates use standardized raw variables measured in the reference simulation ensemble: $\phieff=0.707\,z(\phistar)-0.707\,z(\beta)$ and $\meff=0.181\,z(\phistar)+0.983\,z[\log_{10}(\MBHmax/10^{10}\,\Msun)]$.  The first subtracts the covariance between source normalization and $\beta$.  The second is dominated by the high-mass cutoff and contains a smaller normalization component.  Alternative selection criteria in the evaluation ensemble select the same limiting vector for $\phieff$, and the best-scoring high-mass vector lies within $0.57^\circ$ of the adopted $\meff$ angle.  In the matched three-dimensional comparison, the median precision-axis alignment with $\meff$ is $0.994$.

In the forward model, $\phistar$ controls source abundance, $\MBHmax$ sets the high-mass cutoff, and $b_{\rm BH}$ controls LSS modulation.  The residence factor $T_f(f)$ defines the $\beta$ response, and the high-mass cutoff $P_M(M)$ produces the rare-source response measured by $\meff$.

\subsection{Calibration, numerical resolution, populations, and PTA configurations}
\label{app:model-observing-sensitivity}

\begin{figure*}
    \centering
    \includegraphics[width=\textwidth]{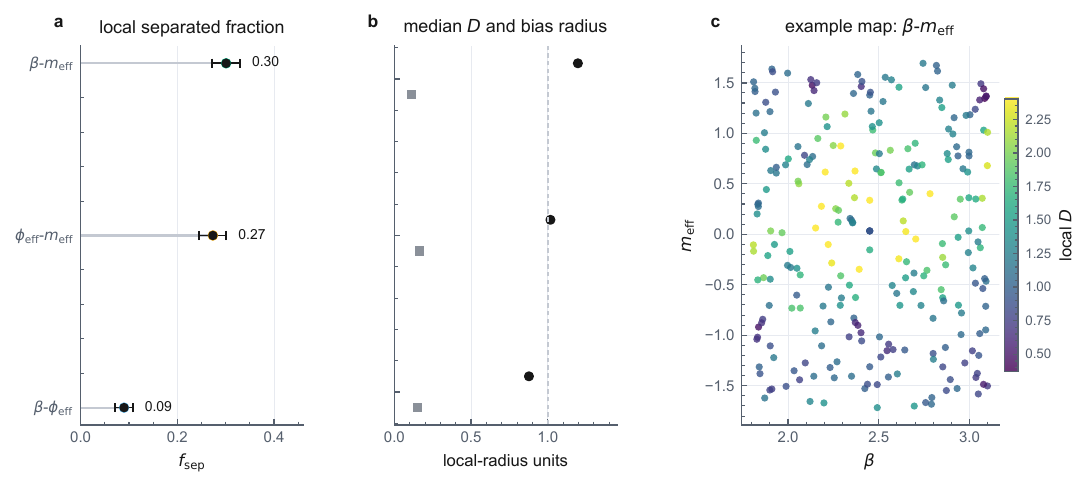}
\caption{Local posterior-radius separation.  The left panel gives the locally separated fraction with binomial standard errors, the middle panel compares the median separation score and bias radius, and the right panel shows the $\beta-\meff$ plane.}
    \label{fig:recoverability-distinguishability}
\end{figure*}

\begin{figure*}
    \centering
    \includegraphics[width=0.94\textwidth]{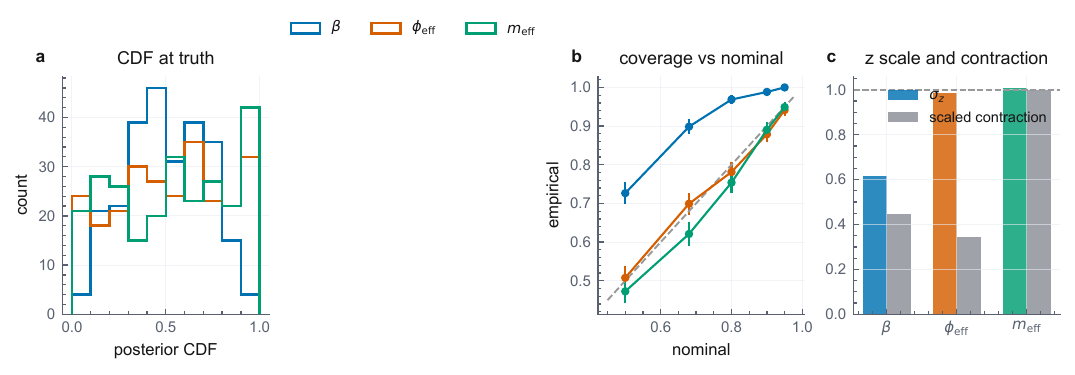}
\caption{Posterior calibration in the evaluation ensemble.  The panels show posterior CDF values at the simulated coordinates, empirical coverage as a function of nominal coverage, and the scale of posterior-centred residuals.}
    \label{fig:posterior-calibration}
\end{figure*}

For the $N=256$ evaluation ensemble, the central 90 per cent coverages are $0.938\pm0.015$ for $\beta$, $0.871\pm0.021$ for $\phieff$, and $0.898\pm0.019$ for $\meff$.  Figure~\ref{fig:posterior-calibration} shows the posterior CDF and empirical coverage diagnostics.  Refining the source grid leaves the response signs unchanged: $\beta$ controls the spectral response, $\phieff$ controls the compensated normalization response, and $\meff$ controls the high-mass contribution.  Figure~\ref{fig:recoverability-distinguishability} summarizes the local posterior-radius separation and posterior-mean offsets for the three coordinate planes.

\begin{figure*}
    \centering
    \includegraphics[width=0.92\textwidth]{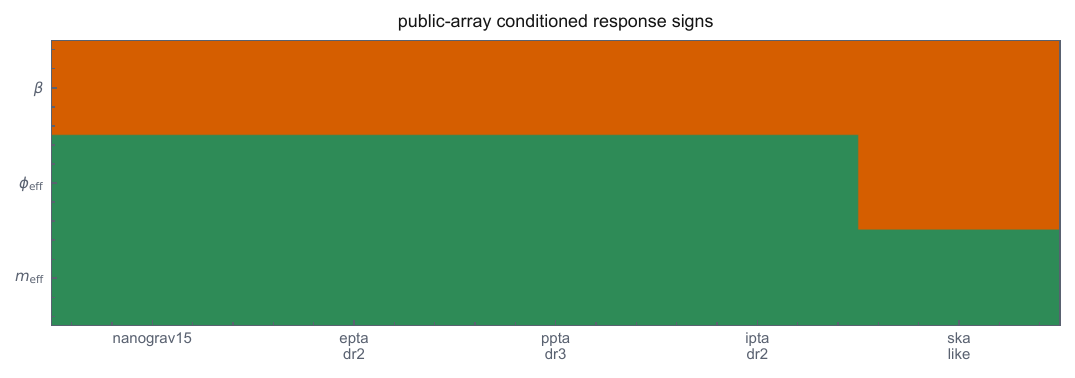}
    \vspace{0.5em}
    \includegraphics[width=0.92\textwidth]{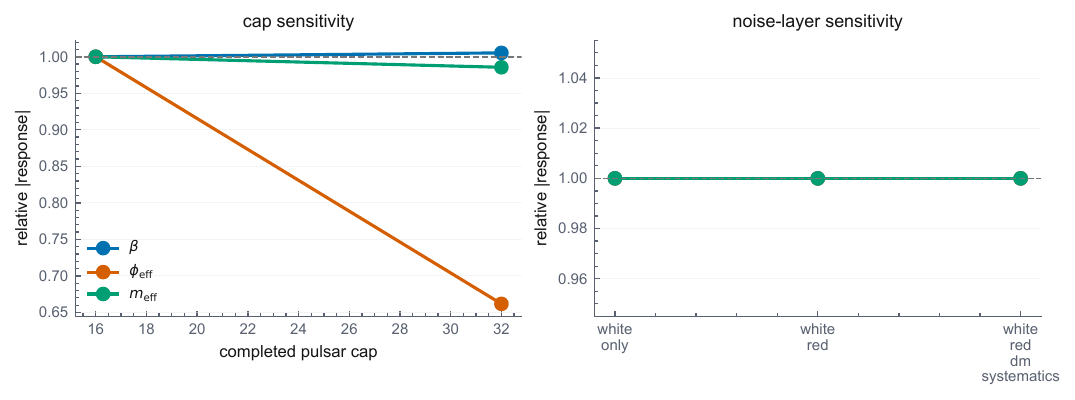}
\caption{Responses for representative PTA configurations.  The panels show the effects of array configuration, array-size caps, and white, red, and dispersion-measure noise.}
    \label{fig:pta-configuration-responses}
\end{figure*}

Figure~\ref{fig:pta-configuration-responses} compares representative NANOGrav, EPTA, PPTA, IPTA, and SKA observing configurations.  The calculations vary sky coverage, array size, white noise, red noise, and dispersion-measure systematics.  The signs of the $\beta$ and $\meff$ responses are unchanged across these configurations.

The population variants change host-relation scatter, the high-mass tail, mass-ratio mapping, eccentricity turnover, environmental evolution, and redshift delay.  The three coordinate-response signs are unchanged across the variants considered.  Posterior calculations for three representative variants use 2048 simulations to construct each posterior and 256 evaluation simulations per variant.

\begin{figure*}
    \centering
    \includegraphics[width=\textwidth]{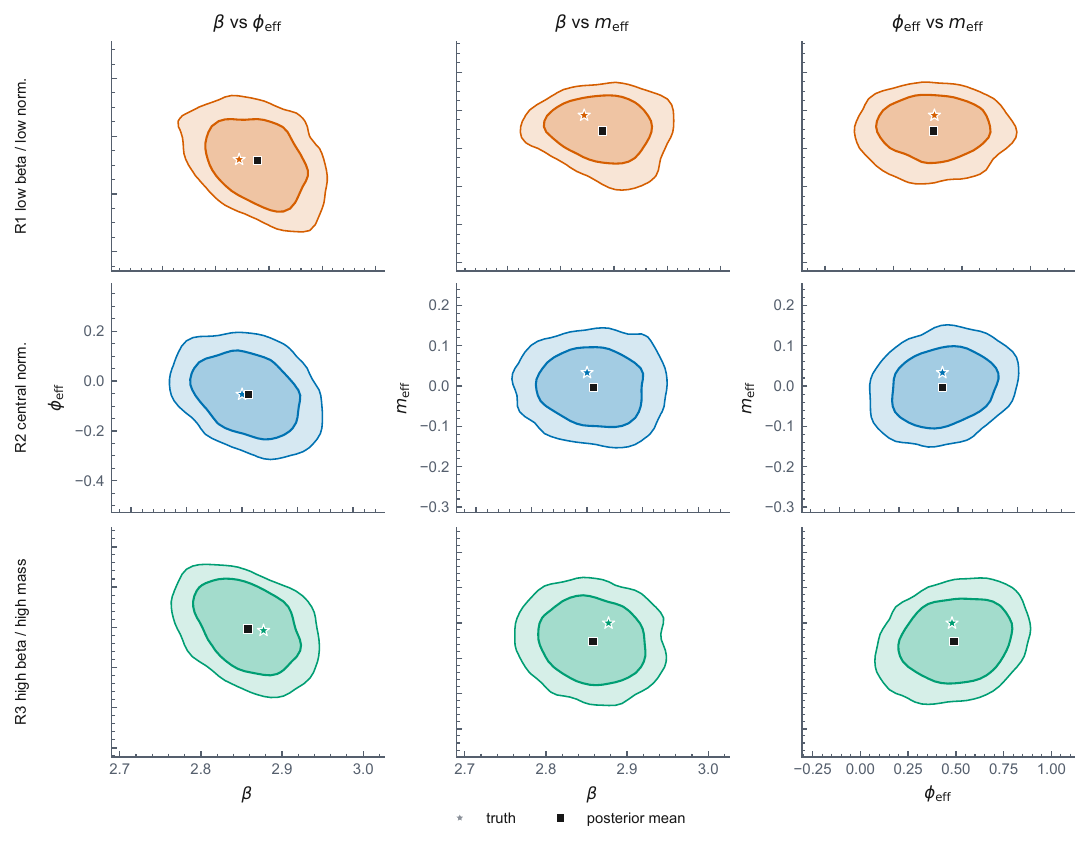}
\caption{Representative joint posteriors.  Stars mark the simulated values and black squares mark posterior means for low-$\beta$/low-normalization, central-normalization, and high-$\beta$/high-mass simulations.}
    \label{fig:representative-posterior}
\end{figure*}

Figure~\ref{fig:representative-posterior} shows joint posteriors for three representative simulations, and Table~\ref{tab:fig5-zscores} lists their posterior-centred $z$-scores.  In each example the simulated values lie within approximately one posterior standard deviation of the posterior means.

\begin{table}
    \centering
    \caption{Posterior-centred $z$-scores for the representative posterior examples in Fig.~\ref{fig:representative-posterior}.}
    \label{tab:fig5-zscores}
    \begin{tabular}{lccc}
        \hline
        Regime & $z_\beta$ & $z_{\phieff}$ & $z_{\meff}$ \\
        \hline
        low-$\beta$/low-$\phieff$ & $-0.531$ & $0.021$ & $0.660$ \\
        central normalization anchor & $-0.173$ & $-0.004$ & $0.550$ \\
        high-$\beta$/high-mass loading & $0.455$ & $-0.052$ & $0.626$ \\
        \hline
    \end{tabular}
\end{table}

\subsection{Matched three-dimensional comparisons}
\label{app:matched-coordinate-comparisons}

The fixed-dimensional comparison uses matched posteriors for $(\beta,\phistar,\ell_M)$ and $(\beta,\phieff,\meff)$, with $\ell_M=\log_{10}(M_{\rm BH,max}/10^{10}M_\odot)$.  Both calculations use the same reference and evaluation simulations and the same posterior-estimation settings.  The median maximum posterior correlation decreases from $0.455$ to $0.383$, and the median $\log_{10}$ covariance condition decreases from $5.68$ to $1.21$.  The fraction with any target outside $2\sigma$ decreases from $0.129$ to $0.098$, and the mean absolute 90 per cent coverage error decreases from $0.059$ to $0.040$.

\begin{table}
    \centering
    \caption{Posterior summaries in matched three-dimensional coordinate systems.}
    \label{tab:selected-basis-gain}
    \resizebox{\columnwidth}{!}{%
    \begin{tabular}{lcc}
        \hline
        Metric & Raw coordinates & Population coordinates \\
        \hline
        max posterior correlation & $0.455$ & $0.383$ \\
        $\log_{10}$ covariance condition & $5.68$ & $1.21$ \\
        any target $|z|>2$ & $0.129$ & $0.098$ \\
        mean $|C_{90}-0.9|$ & $0.059$ & $0.040$ \\
        \hline
    \end{tabular}
    }
\end{table}

\begin{figure*}
    \centering
    \includegraphics[width=0.92\textwidth]{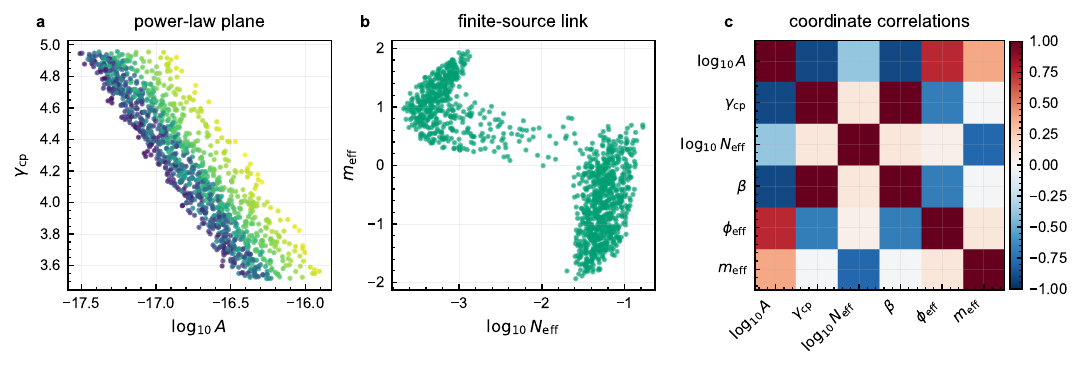}
    \vspace{0.5em}
    \includegraphics[width=0.92\textwidth]{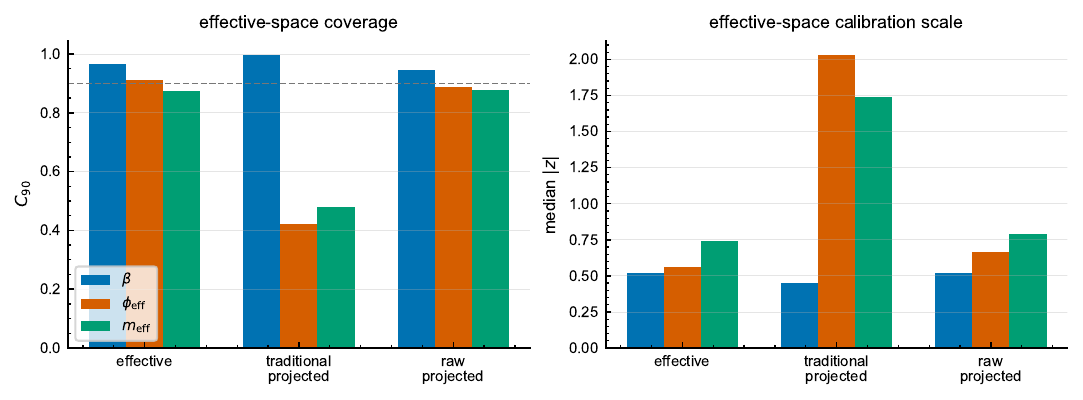}
\caption{Projection from common-process to population coordinates.  The upper panel relates amplitude, spectral index, and source participation to the population coordinates.  The lower panel compares direct inference with the projection of the common-process posterior.}
    \label{fig:common-process-projection}
\end{figure*}

Figure~\ref{fig:common-process-projection} shows how the common-process comparison maps $(\log_{10}A,\gamma_{\rm cp},\log_{10}\Neff)$ into the three population coordinates.  Its native coverages are $0.977$, $0.996$, and $0.922$.  A linear map fitted to the reference simulations projects those posterior samples into $(\beta,\phieff,\meff)$ and is evaluated on the same 256 simulations used for the direct population-coordinate posterior.  The projected $\meff$ coverage is $0.480$, compared with $0.875$ for direct inference, and the direct calculation reduces the absolute error by $0.251$ with a 95 per cent bootstrap interval of $[0.223,0.279]$.

\subsection{Angular response and public free-spectrum comparison}
\label{app:angular-frontier}

\begin{figure*}
    \centering
    \includegraphics[width=0.92\textwidth]{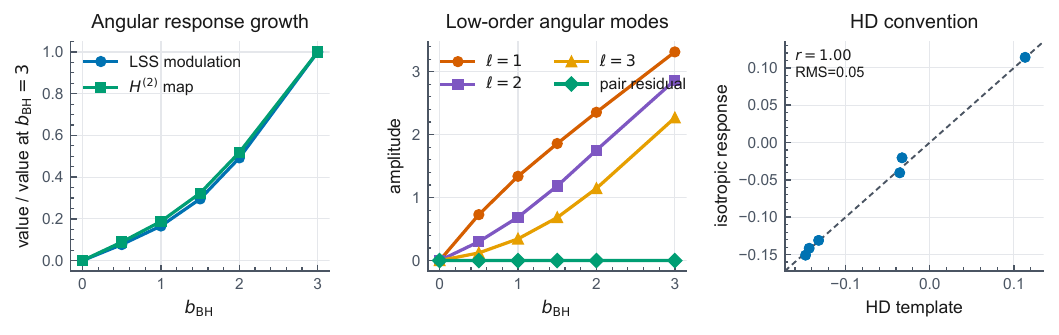}
    \caption{Angular response of the SMBHB population model.  The left and middle panels show the LSS modulation, strain-map anisotropy, and low-order angular response as functions of $b_{\rm BH}$.  The right panel compares the isotropic pair response with the Hellings--Downs curve.}
    \label{fig:angular-frontier}
\end{figure*}

Figure~\ref{fig:angular-frontier} summarizes the angular response at fixed median population parameters for one LSS realization.  Increasing $b_{\rm BH}$ from $0$ to $3$ increases both the LSS modulation RMS and the fractional RMS of the integrated $H^{(2)}$ sky map.  In the isotropic limit, the off-diagonal response has correlation $0.999$ with the Hellings--Downs curve and fractional residual scatter $0.032$.  These calculations characterize the forward angular response.  The posterior target contains the three sky-averaged coordinates in Eq.~\eqref{eq:population-target}.

The NANOGrav 15-year free spectrum supplies an observational comparison for the amplitude and spectral-index responses \citep{NANOGrav:2023gor,NANOGrav:2023hfp,NANOGravKDE15yr:2023,Lamb:2023rapid,Sato-Polito:2024lew,Lamb:2024gbh}.  In the model response, $\beta$ changes the spectral index most directly, $\phieff$ changes the amplitude, and $\meff$ is primarily visible through finite-source information.

\subsection{Illustrative observing-span rescaling}
\label{app:width-rescaling}

\begin{figure*}
    \centering
    \includegraphics[width=0.92\textwidth]{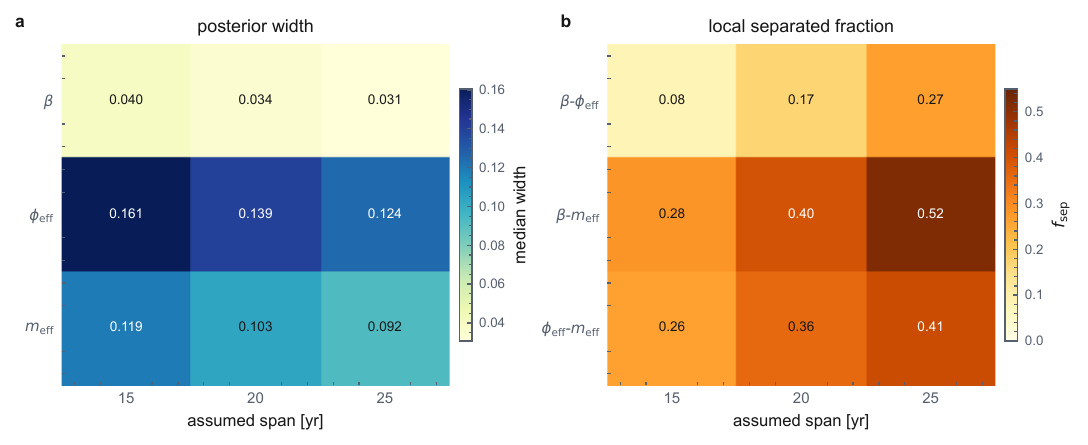}
\caption{Illustrative observing-span rescaling.  The tiles show one-dimensional posterior widths and locally separated fractions under Eq.~\eqref{eq:time-scaling}.}
    \label{fig:width-rescaling}
\end{figure*}

The observation-time exercise rescales posterior widths according to
\begin{equation}
    \sigma(T)
    =
    \sigma(15\,{\rm yr})
    \sqrt{\frac{15\,{\rm yr}}{T}} ,
    \label{eq:time-scaling}
\end{equation}
with the summary definition, population model, cadence, pulsar number, noise assumptions, and prior held fixed.  Figure~\ref{fig:width-rescaling} shows that this illustrative rescaling changes the $\beta-\meff$ locally separated fraction from $0.281$ at 15 yr to $0.398$ at 20 yr and $0.516$ at 25 yr.  The median widths at 25 yr are $0.0308$ for $\beta$, $0.1243$ for $\phieff$, and $0.0920$ for $\meff$.  These values are consequences of the assumed width law.  A PTA forecast requires a new observing simulation.

Across the 1--8 nHz grid, $\dd\gamma_{\rm cp}/\dd\beta=1.000$, consistent with Eq.~\eqref{eq:analytic-gamma-bridge}.  The same calculation gives $\dd\log_{10}A/\dd\phieff=0.267$ and $\dd\gamma_{\rm cp}/\dd\meff\simeq0$.

\clearpage
\makeatletter
\if@filesw
  \immediate\write\@auxout{\string\newlabel{lastpage}{{}{\number\numexpr\value{page}-1\relax}{}{}{}}}%
\fi
\makeatother
\end{document}